\documentclass[prd,superscriptaddress,amsfonts,amssymb,amsmath,showpacs,twocolumn]{revtex4-2}
\usepackage{bm}
\usepackage{amsfonts}
\usepackage{latexsym}
\usepackage[latin1]{inputenc}
\usepackage{graphicx}
\usepackage{amsmath}
\usepackage{palatino}
\usepackage{mathpazo}
\usepackage{textcomp}
\usepackage{float}
\usepackage{booktabs}
\usepackage{dcolumn}
\usepackage{multirow}
\usepackage{ragged2e}
\usepackage{hyperref}
\hypersetup{colorlinks,citecolor=blue}
\usepackage{amsmath}
\usepackage{xcolor}
\usepackage{orcidlink}
\usepackage[caption=false]{subfig}
\usepackage{commath}
\def\jnl@style{\it}
\def\aaref@jnl#1{{\jnl@style#1}}

\def\aaref@jnl#1{{\jnl@style#1}}

\def\aj{\aaref@jnl{AJ}}                   % Astronomical Journal
\def\apj{\aaref@jnl{ApJ}}                 % Astrophysical Journal
\def\apjl{\aaref@jnl{ApJ}}                % Astrophysical Journal, Letters
\def\apjs{\aaref@jnl{ApJS}}               % Astrophysical Journal, Supplement
\def\apss{\aaref@jnl{Ap\&SS}}             % Astrophysics and Space Science
\def\aap{\aaref@jnl{A\&A}}                % Astronomy and Astrophysics
\def\aapr{\aaref@jnl{A\&A~Rev.}}          % Astronomy and Astrophysics Reviews
\def\aaps{\aaref@jnl{A\&AS}}              % Astronomy and Astrophysics, Supplement
\def\mnras{\aaref@jnl{Mon.~Not.~Roy.~Astron.~Soc.}}             % Monthly Notices of the RAS
\def\prd{\aaref@jnl{Phys.~Rev.~D}}        % Physical Review D
\def\prc{\aaref@jnl{Phys.~Rev.~C}}  % Physical Review C
\def\prl{\aaref@jnl{Phys.~Rev.~Lett.}}    % Physical Review Letters
\def\qjras{\aaref@jnl{QJRAS}}             % Quarterly Journal of the RAS
\def\skytel{\aaref@jnl{S\&T}}             % Sky and Telescope
\def\ssr{\aaref@jnl{Space~Sci.~Rev.}}     % Space Science Reviews
\def\zap{\aaref@jnl{ZAp}}                 % Zeitschrift fuer Astrophysik
\def\nat{\aaref@jnl{Nature}}              % Nature
\def\aplett{\aaref@jnl{Astrophys.~Lett.}} % Astrophysics Letters
\def\apspr{\aaref@jnl{Astrophys.~Space~Phys.~Res.}} % Astrophysics Space Physics Research
\def\physrep{\aaref@jnl{Phys.~Rep.}}      % Physics Reports
\def\physscr{\aaref@jnl{Phys.~Scr}}       % Physica Scripta
\def\commat{\aaref@jnl{Comm.~Math.~Phys.}}              % Communications in Mathematical Physics
\def\science{\aaref@jnl{Science}}               % Science
\def\cqg{\aaref@jnl{Classical Quant.~Grav.}}            % Classical and Quantum Gravity
\def\jpcs{\aaref@jnl{JPCS}}                                     % Journal of Physics Conference Series
\def\ijmpd{\aaref@jnl{Int.~J.~Mod.~Phys.~D}}                    % International Journal of Modern Physics D
\def\grg{\aaref@jnl{Gen.~Relat.~Gravit.}}               % General Relativity and Gravitation
\def\rpp{\aaref@jnl{Rep.~Prog.~Phys.}}          % Reports on Progress in Physics
\def\npa{\aaref@jnl{Nucl.~Phys.~A}}        % Nuclear Physics A
\def\lrr{\aaref@jnl{Living Rev.~Rel.}}                   % Living reviews in relativity
\def\jcap{\aaref@jnl{J.~Cosmology Astropart.~Phys.}}    % Journal of cosmology and astroparticle physics
\def\rmp{\aaref@jnl{Rev.~Mod.~Phys.}}   %Reviews of modern physics
\def\epjc{\aaref@jnl{Eur.~Phys.~J.~C}}

\allowdisplaybreaks[1]
\begin{document}

\color{black}       %% For one column
\title{Extended Bose-Einstein condensate dark matter in $f(Q)$ gravity}

\author{Aaqid Bhat\orcidlink{xxxx}}
\email{aaqid555@gmail.com}
\affiliation{Department of Mathematics, Birla Institute of Technology and
Science-Pilani,\\ Hyderabad Campus, Hyderabad-500078, India.}
\author{Raja Solanki\orcidlink{0000-0001-8849-7688}}
\email{rajasolanki8268@gmail.com}
\affiliation{Department of Mathematics, Birla Institute of Technology and
Science-Pilani,\\ Hyderabad Campus, Hyderabad-500078, India.}

\author{P.K. Sahoo\orcidlink{0000-0003-2130-8832}}
\email{pksahoo@hyderabad.bits-pilani.ac.in}
\affiliation{Department of Mathematics, Birla Institute of Technology and
Science-Pilani,\\ Hyderabad Campus, Hyderabad-500078, India.}
%%%%%%%%%%%%%%%%%%%%%%%%%%%%%%%%%%%%  DATE  %%%%%%%%%%%%%%%%%%%%%%%%%%%%%%%%%%%%

\date{\today}
\begin{abstract}
In this article, we attempt to explore the dark sector of the universe i.e. dark matter and dark energy, where the dark energy components are related to the modified $f(Q)$ Lagrangian, particularly a power law function $f(Q)= \gamma \left(\frac{Q}{Q_0}\right)^n$, while the dark matter component is described by the Extended Bose-Einstein Condensate (EBEC) equation of state for dark matter, specifically, $p = \alpha \rho + \beta \rho^2$. We find the corresponding Friedmann-like equations and the continuity equation for both dark components along with an interacting term, specifically $\mathcal{Q} = 3b^2H \rho$, which signifies the energy exchange between the dark sector of the universe. Further, we derive the analytical expression of the Hubble function, and then we find the best-fit values of free parameters utilizing the Bayesian analysis to estimate the posterior probability and the Markov Chain Monte Carlo (MCMC) sampling technique corresponding to CC+Pantheon+SH0ES samples. In addition, to examine the robustness of our MCMC analysis, we perform a statistical assessment using the Akaike Information Criterion (AIC) and Bayesian Information Criterion (BIC). Further from the evolutionary profile of the deceleration parameter and the energy density, we obtain a transition from the decelerated epoch to the accelerated expansion phase, with the present deceleration parameter value as $q(z=0)=q_0=-0.56^{+0.04}_{-0.03}$ ($68 \%$ confidence limit), that is quite consistent with cosmological observations. In addition, we find the expected positive behavior of the effective energy density. Finally, by examining the sound speed parameter, we find that the assumed theoretical $f(Q)$ model is thermodynamically stable. 
 
\end{abstract}

\maketitle

\textbf{Keywords:} $f(Q)$ gravity, Bose-Einstein Condensate, dark matter, and dark energy.

\section{Introduction}\label{sec1}

As is widely recognized, the universe is comprised of both visible and dark components. The visible elements encompass all visible entities within the cosmos, while the dark components encompass dark matter and dark energy. Among these mysterious components, dark matter stands out as an undetectable entity within the electromagnetic radiation spectrum. The observational phenomena such as cosmic microwave background and the gravitational lensing \cite{R1,R2,R3,R4} provide compelling evidence for the presence of dark matter, aiming to elucidate the difference between the estimated mass of large celestial bodies and the mass derived from luminous matter like stars, gas, and dust within them. This implies that empirical data indicates the emergence of dark matter through gravitational pull on ordinary matter. The dark matter encompasses both baryonic and non-baryonic forms. In the baryonic form, dark matter manifests as astronomical entities like massive and compact haloes, primarily made up of ordinary baryonic matter yet emitting negligible electromagnetic radiation. Conversely, non-baryonic dark matter is characterized by hypothetical and actual particles, whereas the Weakly Interacting Massive Particles and axions are the hypothetical ones.  Additionally, the state of matter known as Bose-Einstein condensate (BEC) arises in the non-baryonic realm, formed when particles called bosons undergo cooling to near absolute zero \cite{R5,R6,R7,R8,R9,R10,R11}. As a consequence of the extremely low temperature, a phase transition takes place, causing the majority of boson gases to occupy the lowest quantum state, leading to the manifestation of macroscopic quantum phenomena. At this condition, cold bosons interact, giving rise to superparticles that exhibit microwave-like behavior \cite{R12}. The assumption is made that dark matter exists in the form of a bosonic gas below a critical temperature, leading to the formation of Bose-Einstein condensate (BEC). For more on BEC, see the references \cite{ADD1,ADD2,ADD3,ADD4,ADD5}. Utilizing the generalized Gross-Pitaevskii equation, the equation of state (EoS) for dark matter is derived as that of a barotropic fluid. It is noteworthy that this particular EoS is referred to as conventional dark matter. Through the consideration of the dark matter halo existing in a quantum ground state, the equation of state (EoS) was derived as $p \propto \rho^2$ \cite{R10}.  It is important to highlight that the origins of both usual dark matter and the quantum ground state can be attributed to the one-body and two-body interactions among bosonic particles. Note that, the equation of state (EoS) $p=0$, $p=\alpha \rho$, and $p=\beta \rho^2$ characterize the cold dark matter, normal dark matter, and dark matter halo respectively. These observations prompt the introduction of the Extended Bose-Einstein Condensate (EBEC) model, a comprehensive model combining normal dark matter and the quantum ground state \cite{R13}. The merit of this approach lies in its capacity to concurrently account for both one-body and two-body interactions, offering insights into the components of the universe, particularly dark matter.

Further, the conventional theory of relativity, particularly General Relativity, which interprets gravity as the curvature of spacetime, may not provide the ultimate solution for explaining dark energy. This encourages the exploration of alternative theoretical frameworks in cosmology that can effectively account for cosmic acceleration while aligning with observational data. General Relativity and its curvature-based extensions have been formulated and adequately studied in the past \cite{R14,R15}. Recently, modified theories of gravitation within a flat spacetime geometry, dependent solely on non-metricity have been established and extensively investigated \cite{R16,R17}. Various astrophysical and cosmological implications of the $f(Q)$ gravity have been widely investigated \cite{R18,R19,R20,R21,R22,R23,R24,R25,R26,R27,R28,R29}. The manuscript is organized as follows: In section \eqref{sec2}, we present the mathematical background of the $f(Q)$ gravity. In section \eqref{sec3}, we employ EBEC dark matter EoS along with the power law $f(Q)$ lagrangian to derive the analytical solution of the Friedmann-like equations. Further in section \eqref{sec4}, we find the best-fit values of free parameters utilizing the Bayesian analysis to estimate the posterior probability and the Markov Chain Monte Carlo (MCMC) sampling. In addition, we employ AIC and BIC tools to examine the robustness of our MCMC analysis and then we test the stability of the considered cosmological $f(Q)$ model. Finally in section \eqref{sec5}, we discuss our findings.

\section{The Mathematical Formulation of $f(Q)$ Theory}\label{sec2}
\justifying
The standard general relativity is based on the Lorentzian geometry equipped with a metric-compatible symmetric connection known as the Levi-Civita connection. This metric-compatible connection leads to the non-zero curvature with vanishing torsion \cite{R290}. Further, one can utilize the metric-compatible Weitzenbock connection possessing non-vanishing torsion to establish the teleparallel equivalent to general relativity (TEGR). However, the most generic metric-affine connection cab be expressed as follows,
\begin{equation}\label{2a}
\hat{\Gamma}^{\,\sigma}_{\,\,\,\alpha\beta}=\Gamma^{\,\sigma}_{\,\,\,\alpha\beta}+K^{\,\sigma}_{\,\,\,\alpha\beta}+L^{\,\sigma}_{\,\,\,\alpha\beta}
\end{equation}
where the first quantity represents the aforementioned Levi-Civita connection and can be expressed as $\Gamma^{\,\sigma}_{\,\,\,\alpha\beta}=\frac{1}{2}g^{\sigma\lambda}\left(\partial_{\alpha}g_{\lambda\beta}+\partial_{\beta}g_{\lambda\alpha}-\partial_{\lambda}g_{\alpha\beta}\right)$ , the second one is the contorsion tensor and can be estimated as $K^{\,\sigma}_{\,\,\,\alpha\beta}=\frac{1}{2}T^{\,\sigma}_{\,\,\,\alpha\beta}+T^{\,\,\,\,\,\,\,\sigma}_{(\alpha\,\,\,\,\,\,\beta)}$, and the last one $L^{\,\sigma}_{\,\,\,\alpha\beta}=-\frac{1}{2}g^{\sigma\lambda}\left(Q_{\alpha\lambda\beta}+Q_{\beta\lambda\alpha}-Q_{\lambda\alpha\beta}\right)$ is the disformation tensor.\\
One can establish the symmetric teleparallel equivalent to GR (STEGR) utilizing the symmetric connection satisfying the non-metricity condition with vanishing torsion, and hence the gravitational interaction is attributed to the obtained non-metricity tensor that can be expressed as,
\begin{equation}\label{2b}
Q_{\sigma\alpha\beta}=\nabla_{\sigma}g_{\alpha\beta},
\end{equation}
and the corresponding traces are
\begin{equation}\label{2c}
Q_{\sigma}=Q_{\sigma\,\,\,\,\alpha}^{\,\,\,\,\alpha}\, ,\,\,\,\,\,\,\,\,\tilde{Q}_{\sigma}=Q^{\alpha}_{\,\,\,\,\sigma\alpha}\,.
\end{equation}
In addition, its conjugate is given by,
\begin{equation}\label{2d}
4P_{\,\,\alpha\beta}^{\sigma}=-Q^{\sigma}_{\,\,\,\,\alpha\beta}+2Q^{\,\,\,\,\,\,\sigma}_{(\alpha\,\,\,\,\beta)}-Q^{\sigma}g_{\alpha\beta}-\tilde{Q}^{\sigma}g_{\alpha\beta}-\delta^{\sigma}_{(\alpha}\, Q\,_{\beta)},
\end{equation}
We define the non-metricity scalar as,
\begin{equation}\label{2e}
Q=-Q_{\sigma\alpha\beta}P^{\sigma\alpha\beta}.
\end{equation}
The generic action governing the $f(Q)$ gravity along with Lagrange multiplier expressed as \cite{R21},
\begin{equation}\label{2f}
S=\int \left[\frac{1}{2}f(Q)+\lambda_{\alpha}^{\,\,\,\beta\mu\nu} R^{\alpha}_{\,\,\,\beta\mu\nu}+\lambda_{\alpha}^{\,\,\,\mu\nu} T^{\alpha}_{\,\,\,\mu\nu}+\mathcal{L}_m\right]\sqrt{-g}\,d^4x,
\end{equation}
Here $f(Q)$ is the function of non-metricity scalar term, $g=det(g_{\alpha\beta})$, $\mathcal{L}_m$ denotes the Lagrangian density of matter, and $\lambda_{\alpha}^{\,\,\,\beta\mu\nu}$ is the Lagrange multiplier. In particular, for the functional choice $f(Q) = -Q$, we retrieve the STEGR case.
Note that the geometrical settings presented here, utilizes a flat torsion-free connection that corresponds to a coordinate transformation from the trivial connection as prescribed in \cite{R17}. Moreover it can be parameterized using $\xi^{\alpha}$, as follows,
\begin{equation}\label{2g}
\hat{\Gamma}^{\,\sigma}_{\,\,\,\alpha\beta}=\frac{\partial x^{\sigma}}{\partial\xi^{\mu}}\partial_{\alpha}\partial_{\beta}\xi^{\mu}
\end{equation}
Also note that $\xi^{\alpha}=\xi^{\alpha}(x^{\sigma})$ is an invertible relation. Now, one can always choose a transformation in such a way that the generic affine connection vanishes (i.e., $\hat{\Gamma}^{\,\sigma}_{\,\,\,\alpha\beta}=0$), such a coordinate transformation is known as the coincident gauge. Consequently, we obtained the non-metricity component as $Q_{\sigma\alpha\beta}=\partial_{\sigma}g_{\alpha\beta}$.

Now, the energy-momentum tensor reads as
\begin{equation}\label{2h}
T_{\alpha\beta}\equiv-\frac{2}{\sqrt{-g}}\frac{\delta(\sqrt{-g}\mathcal{L}_m)} {\delta g^{\alpha\beta}}.
\end{equation}
and the field equation of the $f(Q)$ gravity obtained by varying the action \eqref{2f} with respect to the metric, within a symmetric teleparallelism framework, given as
\begin{small}
\begin{equation}\label{2i}
\frac{2}{\sqrt{-g}}\nabla_{\sigma}\left(f_{Q}\sqrt{-g}\,P^{\sigma}_{\,\,\alpha\beta}\right)+\frac{1}{2}f\,g_{\alpha\beta}+
f_{Q}\left(P_{\alpha\sigma\lambda}Q_{\beta}^{\,\,\,\sigma\lambda}-2Q_{\sigma\lambda\alpha}P^{\sigma\lambda}_{\,\,\,\,\,\,\beta}\right)=- T_{\alpha\beta},
\end{equation}
\end{small}
where $f_Q=\frac{d f}{d Q}$. In addition, on varying the action \eqref{2f} with respect to the connection yields,
\begin{equation}\label{2j}
\nabla_{\sigma}\lambda_{\mu}^{\,\,\,\alpha\beta\sigma}+\lambda_{\mu}^{\,\,\,\alpha\beta}=\sqrt{-g}f_Q\,P_{\,\mu}^{\,\,\,\alpha\beta}+H_{\,\mu}^{\,\,\,\alpha\beta}
\end{equation}
where $H_{\,\mu}^{\,\,\,\alpha\beta}=-\frac{1}{2}\frac{\delta\mathcal{L}_m}{\delta\Gamma^{\mu}_{\,\,\,\alpha\beta}}$ is the hypermomentum tensor density. Now,
by using the antisymmetry property of $\alpha$ and $\beta$ in the Lagrangian multiplier coefficients, the above equation becomes,
\begin{equation}\label{2k}
\nabla_{\alpha}\nabla_{\beta} \left(f_{Q}\sqrt{-g}\,P_{\,\mu}^{\,\,\,\alpha\beta}+H_{\,\mu}^{\,\,\,\alpha\beta}\right)=0.
\end{equation}
Note that the variation of the connection with respect to $\xi^{\sigma}$ is equivalent to performing a diffeomorphism such that $\delta_{\xi}\hat{\Gamma}^{\,\sigma}_{\,\,\,\alpha\beta}=-\mathcal{L}_{\xi}\hat{\Gamma}^{\,\sigma}_{\,\,\,\alpha\beta}=-\nabla_{\alpha}\nabla_{\beta}\xi^{\sigma}$, where the utilized connection is flat and torsion-free \cite{R20}. In absence of the hypermomentum, the above equations becomes \cite{R21},
\begin{equation}\label{2l}
\nabla_{\alpha}\nabla_{\beta} \left(f_{Q}\sqrt{-g}\,P_{\,\mu}^{\,\,\,\alpha\beta}\right)=0.
\end{equation}
For both connection and metric field equations, one can acquire that $\mathcal{D}_{\alpha}T^{\alpha}_{\,\,\,\,\beta}=0$, for the metric-covariant derivative $\mathcal{D}_{\alpha}$. However, in the presence of hypermomentum, a relationship between the hypermomentum density and the divergence of the energy-momentum tensor can be obtained \cite{R291}.\\
One can obtained the energy-momentum tensor for the perfect fluid type matter distribution as,
\begin{equation}\label{2m}
T_{\alpha\beta}=(p+\rho)u_{\alpha}u_{\beta}+pg_{\alpha\beta},
\end{equation}
where $p$ and $\rho$ denotes the usual pressure component and the energy density respectively, along with $u_{\alpha}$ as four-velocity vector. \\
We assume the following homogeneous and isotropic flat FLRW line element,
\begin{equation}\label{2n}
ds^2=-dt^2+a^2(t)(dx^2+dy^2+dz^2),
\end{equation}
where $a(t)$ is the scale factor. We obtained the corresponding non-metricity scalar as $Q=6H^2$. The Friedmann like equations for the generic $f(Q)$ functional corresponding to the line element \eqref{2n} obtained as follows \cite{R292},
\begin{equation}\label{2o}
3H^2=\frac{1}{2f_Q} \left( -\rho+\frac{f}{2}  \right)
\end{equation}
\begin{equation}\label{2p}
\dot{H}+3H^2+ \frac{\dot{f_Q}}{f_Q}H = \frac{1}{2f_Q} \left( p+\frac{f}{2} \right)
\end{equation}
We can rewrite equations \eqref{2o} and \eqref{2p} as follows, 
\begin{equation}\label{2q}
3H^2= \rho + \rho_{de}    
\end{equation}
\begin{equation}\label{2r}
 \dot{H}=- \frac{1}{2}\left[  \rho + \rho_{de} +p + p_{de} \right]   
\end{equation}
where $\rho_{de}$ and $p_{de}$ are energy density and pressure of the dark energy fluid part arising due to non-metricity component, and can be expressed as follows,
\begin{equation}\label{2s}
\rho_{de}= \frac{1}{2} (Q-f) + Q f_Q
\end{equation}
and 
\begin{equation}\label{2t}
 p_{de}=-\rho_{de} - 2\dot{H} (1+f_Q+2Qf_{QQ}) 
\end{equation}
Further, we write the continuity equation for both matter and dark energy component as,
\begin{equation}\label{2u}
\dot{\rho} + 3H(\rho + p) = \mathcal{Q}   
\end{equation}
and
\begin{equation}\label{2v}
\dot{\rho}_{de} + 3H(\rho_{de} + p_{de}) = - \mathcal{Q}   
\end{equation}
where $\mathcal{Q}$ is defined as an interaction term arising due to the energy transfer between dark components of the universe. It is evident that the parameter $\mathcal{Q}$ must possess a positive value, indicating the occurrence of energy transfer from dark energy to dark matter, ensuring the second law of thermodynamics. In this context, considering $\mathcal{Q}$ as the product of the energy density and the Hubble parameter is a natural choice, given that it represents the inverse of cosmic time. Therefore, we adopt the specific expression $\mathcal{Q} = 3b^2H \rho$, where $b$ is the intensity of energy transfer \cite{R13}.\\

\section{DARK MATTER AS A EBEC}\label{sec3}
\justifying
In this section, we examine the normal dark matter as bosonic particles whose number density is determined by Bose-Einstein statistics, signifying the formation of these particles through the decoupling of the residual plasma during the early universe. However, the energy density of dark matter is expressed as the product of particle number density and the mass of dark matter. The pressure of dark matter, adhering to Bose-Einstein statistics, is defined within a sphere characterized by the radius of the particles' momentum \cite{R30,R31}. Consequently, one can express the normal dark matter pressure as a linear relationship in terms of the energy density as follows,
\begin{equation}\label{3a}
p=\alpha \rho    
\end{equation}
Now, we examine the BEC dark matter as non-relativistic bosons engaged in a two-particle interaction within a quantum system. As discussed earlier, BEC recognized as a state of matter, emerges when a dilute Bose gas undergoes significant cooling to attain extremely low temperatures.
According to experimental observations of the Einstein and Bose, as the temperature approaches absolute zero, the waves associated with the particles eventually overlap. This phenomenon results in the merging of elementary particles into a single quantum state, termed Bose-Einstein condensation. The generalized Gross-Pitaevskii equation characterizes the physical behavior of the BEC \cite{R32}. When considering BEC within a gravitational context, the corresponding equation of state (EoS) for BEC dark matter is formulated in the subsequent manner,
\begin{equation}\label{3b}
p=\beta \rho^2    
\end{equation}
where $\beta$ is defined as the coefficient related to the dark matter mass and its scattering length \cite{R33,R34}.
Now in order to gain a more profound comprehension of the universe, we assume an extended form of the equation of state for dark matter known as the Extended Bose-Einstein Condensation (EBEC) for dark matter EoS as \cite{R13}
\begin{equation}\label{3c}
p=\alpha \rho + \beta \rho^2    
\end{equation}
Here, $\alpha$ represent the single-body interaction arising from conventional dark matter, while $\beta$ is introduced to signify the two-body interaction originating from the dark matter halo. In particular, $\alpha=\beta=0$ reduces to the cold dark matter case, whereas the case $\beta=0$ reduces to the normal matter scenario. Further, the case $\alpha=0$ represents dark matter halo, while $\alpha \neq 0$ and $\beta \neq 0$ represents the contribution from both dark matter halo and the normal matter.

We consider the following dynamically tested power-law $f(Q)$ function that can efficiently describes an evolution of the universe from a matter dominated phase to the de-Sitter era \cite{R341},
\begin{equation}\label{3d}
f(Q)= \gamma \bigg(\frac{Q}{Q_0}\bigg)^n
\end{equation}
where $Q_0=6H_0^2$ and $\gamma$ and $n$ are free parameters. Then by using equation \eqref{3d} in the equation \eqref{2q}, we obtained
\begin{equation}\label{3e}
\rho = \frac{(1-2n)}{2}\gamma \left( \frac{H}{H_0} \right)^{2n}  
\end{equation}
On evaluating the equation \eqref{3e} at present redshift $z = 0$, we have
\begin{equation}\label{3f}
\rho_0 = \frac{(1-2n)}{2}\gamma     
\end{equation}
and therefore, we have
\begin{equation}\label{3g}
\rho = \rho_0 \left( \frac{H}{H_0} \right)^{2n}    
\end{equation}
Now, on integrating the continuity equation \eqref{2u} for the matter component, we acquired
\begin{equation}\label{3h}
 \rho = \rho_0 \left( \frac{c \eta - \beta}{c \eta (1+z)^{3 \eta} - \beta}  \right)
\end{equation}
Here $c$ is the constant of integration and $\eta=\alpha+1-b^2$. We obtained the expression of the Hubble parameter, by utilizing equations \eqref{3g} and \eqref{3h}, as follows,
\begin{equation}\label{3i}
H(z)=H_0 \left( \frac{c \eta - \beta}{c \eta (1+z)^{-3 \eta} - \beta}  \right)^{\frac{1}{2n}}
\end{equation}

\section{BEST FIT VALUE OF PARAMETERS}\label{sec4}
\justifying
In this segment, a statistical analysis is conducted to compare the predictions of the theoretical model under consideration with observational data. The aim is to establish constraints on the unrestricted variables within the model. The analysis utilizes a sample of Cosmic Chronometers, comprising 31 measurements, and the Pantheon+SH0ES sample, which includes 1701 data points. Bayesian statistical techniques are employed to estimate the posterior probability through the utilization of the likelihood function and the Markov Chain Monte Carlo (MCMC) random sampling technique.

\subsection{Cosmic Chronometers}
\justifying
Cosmic chronometers are characterized as a collection of predominantly elderly, inactive galaxies that have concluded their star-forming activities, identifiable by specific features in their spectra and color profiles \cite{R35}. The data employed in cosmic chronometers is estimated utilizing the age of these galaxies having different redshifts. In our analysis, we utilized a compilation of 31 independent measurement of $H(z)$ within the redshift range $ 0.07 \leq z \leq 2.41 $ \cite{R36}. These $H(z)$ measurements are derived using the relationship $H(z)=-\frac{1}{1+z} \frac{dz}{dt}$. Here, $\frac{dz}{dt}$ estimated utilizing $\frac{\Delta z}{\Delta t}$, where $\Delta z$ and $\Delta t$ being the variation in redshift and age between two galaxies.  The corresponding $\chi^2$ function is defined as, 
\begin{equation}\label{4a}
\chi_{CC}^{2}=\sum\limits_{k=1}^{31}\frac{[H_{th}(z_{k})-H_{obs,k}]^{2}}{\sigma _{H,k}^{2}}.  
\end{equation}
Here, $H_{obs,k}$ denotes the observed Hubble value and $H_{th}(z_{k})$ denotes the theoretical value of $H(z)$ corresponding to the redshift $z_{k}$ with $\sigma_{H,k}$ as the standard error.

\subsection{Pantheon+SH0ES}
\justifying
The Pantheon+SH0ES samples cover a broad spectrum of redshifts, ranging from 0.001 to 2.3, surpassing previous collections of Type Ia supernovae (SNIa) by incorporating the latest observational data. Type Ia supernovae, renowned for their consistent brightness, serve as reliable standard candles for gauging relative distances utilizing the distance modulus technique. In the last two decades, several compilations of Type Ia supernova data have been introduced, such as Union \cite{R37}, Union2 \cite{R38}, Union2.1 \cite{R39}, JLA \cite{R40}, Pantheon \cite{R41}, and the most recent addition, Pantheon+SH0ES \cite{R42}. The corresponding $\chi^2$ function is expressed as, 
\begin{equation}\label{4b}
\chi^2_{SN}= D^T C^{-1}_{SN} D,
\end{equation}
Here, $C_{SN}$ \cite{R42} represents the covariance matrix associated with the Pantheon+SH0ES samples, encompassing both statistical and systematic uncertainties. Moreover, the vector $D$ is defined as $D=m_{Bi}-M-\mu^{th}(z_i)$, where $m_{Bi}$ and $M$ are the apparent magnitude and absolute magnitude, respectively. In addition, the $\mu^{th}(z_i)$ represents the distance modulus of the assumed theoretical model, and it can be expressed as,
\begin{equation}\label{4c}
\mu^{th}(z_i)= 5log_{10} \left[ \frac{D_{L}(z_i)}{1 Mpc}  \right]+25, 
\end{equation}
where, $D_{L}(z)$ is the luminosity distance assumed theoretical model, and it can be expressed as,
\begin{equation}\label{4d}
D_{L}(z)= c(1+z) \int_{0}^{z} \frac{ dx}{H(x,\theta)}
\end{equation}
where, $\theta$ is the parameter space of the assumed model.

Unlike the Pantheon dataset, the Pantheon+SH0ES compilation effectively addresses the degeneracy between the parameters $H_0$ and $M$ by redefining the vector $D$ as
\begin{equation}\label{4e}
\bar{D} = \begin{cases}
     m_{Bi}-M-\mu_i^{Ceph} & i \in \text{Cepheid hosts} \\
     m_{Bi}-M-\mu^{th}(z_i) & \text{otherwise}
    \end{cases}   
\end{equation}
Here $\mu_i^{Ceph}$ independently estimated using Cepheid calibrators. Hence, the equation \eqref{4b} obtained as $\chi^2_{SN}=  \bar{D}^T C^{-1}_{SN} \bar{D} $.\\

We obtain the constraints on free parameter space for the combined CC+Pantheon+SH0ES samples utilizing the Gaussian priors as $[50,100]$ for $H_0$, $[-5,0]$ for $n$, $[0,5]$ for $\beta$, $[-5,0]$ for $\eta$, and $[0,1]$ for $c$. In order to obtain the best fit value of parameters, we minimize the total $\chi^2_{total}$ function that is defined as follows,
\begin{equation}\label{4f}
\chi^2_{total}= \chi^2_{CC}+\chi^2_{SN} 
\end{equation}
The corresponding contour plot describing the correlation between different model parameters within the $1\sigma-3\sigma$ confidence interval is presented in the Figure \eqref{f1}.
\begin{widetext}

\centering
\begin{figure}[H]
\includegraphics[scale=0.75]{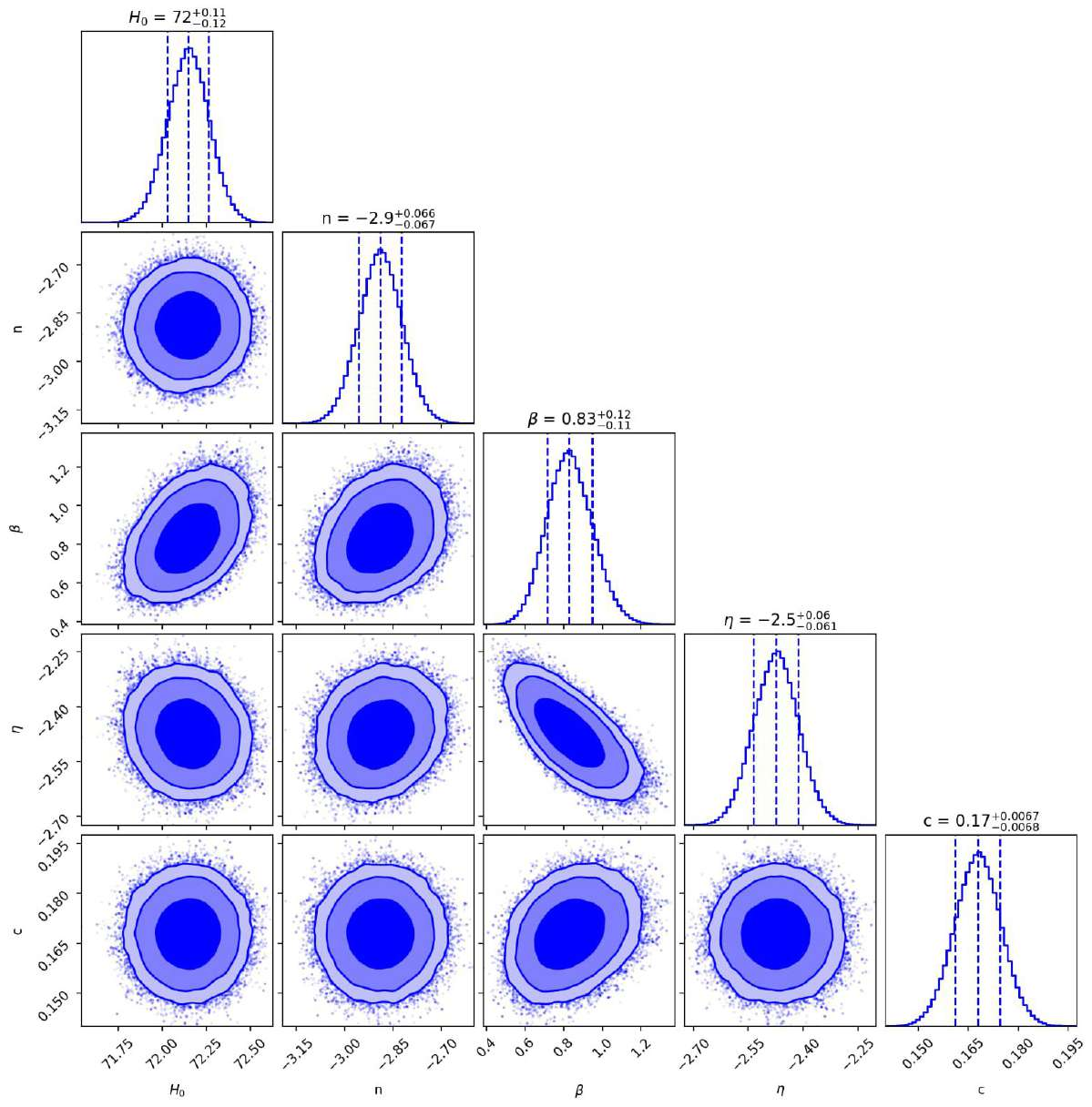}
\caption{The contour plot for the given model corresponding to the free parameter space $(H_0, n, \beta, \eta, c)$ within the $1\sigma-3\sigma$ confidence interval using CC+Pantheon+SH0ES samples.}\label{f1}

\end{figure}
\end{widetext}  
We obtained constraints on the free parameter space with $68 \%$ confidence limit as $H_0=72^{+0.11}_{-0.12}$, $n=-2.9^{+0.066}_{-0.067}$, $\beta=0.83^{+0.12}_{-0.11}$, $\eta=-2.5^{+0.06}_{-0.061}$, and $c=0.17^{+0067}_{-0.0068}$. In addition, we obtained the minimum value of the $\chi^2_{total}$ as $\chi^2_{min}=1642.55$.

\subsection{Model Comparison}
\justifying
To evaluate the robustness of our MCMC analysis, it is crucial to perform a statistical assessment using the Akaike Information Criterion (AIC) and Bayesian Information Criterion (BIC) \cite{R43}. The initial criterion, AIC, can be expressed as follows,
\begin{equation}\label{4g}
AIC = \chi^2_{min} + 2d    
\end{equation}
Here, $d$ represents the number of parameters within the given model. For the model comparison with the established $\Lambda$CDM model, we introduce $\Delta AIC = |AIC_{Model} - AIC_{\Lambda CDM}|$. A value of $\Delta AIC$ less than 2 suggests strong evidence in favor of the assumed theoretical model, while in the range of $4 < \Delta AIC \leq 7$, there is a moderate support. Moreover, if the $\Delta AIC$ value exceeds 10, there is no evidence supporting the assumed model. The second criterion, BIC, can be expressed as follows,
\begin{equation}\label{4h}
BIC = \chi^2_{min} + d ln(N)    
\end{equation}
Here, $N$ represents the number of data samples used in the MCMC analysis. Similarly, $\Delta BIC$ is less than 2 suggests strong evidence in favor of the assumed theoretical model, while in the range of $2 < \Delta BIC \leq 6$, there is a moderate support. Utilizing the aforementioned $\chi^2_{min}$ minimum value, we obtained $AIC_{Model}=1652.55$ and $BIC_{Model}=1679.8$ and hence we obtained $\Delta AIC=0.85$ and $\Delta BIC=15.5$, where $\Lambda$CDM value taken to be $AIC_{\Lambda CDM}=1679.4$ and  $BIC_{\Lambda CDM}=1664.3$. Thus, it is evident from the $\Delta AIC$ value that there is strong evidence in favour of the assumed theoretical $f(Q)$ model. However, it is well known that large number of parameters compensate the high $\Delta BIC $ value.

\subsection{Evolutionary Parameters}
\justifying
The deceleration parameter is an essential tool to quantify the evolutionary phase of expansion of the universe. It is defined as follows,
\begin{equation}\label{4i}
 q=-1-\frac{\dot{H}}{H^2}   
\end{equation}

\begin{figure}[H]
\centering
\includegraphics[scale=0.475]{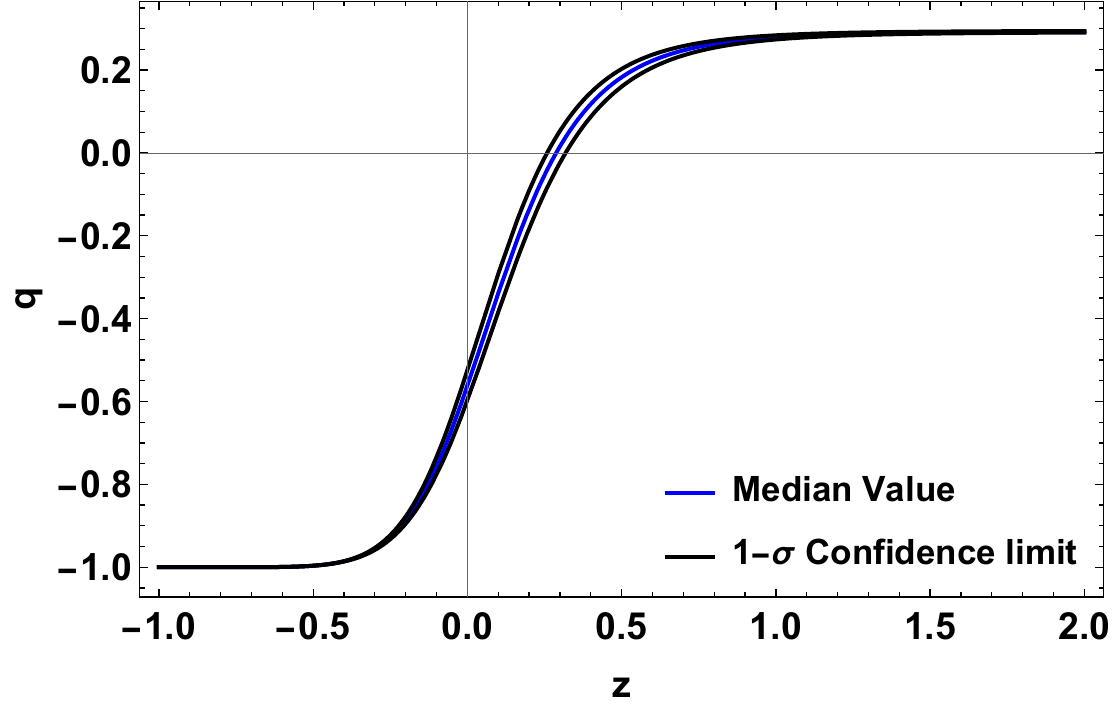}
\caption{Profile of the deceleration parameter vs redshift corresponding to obtained parameter constraints with $68 \%$ confidence limit. }
\label{f2}
\end{figure}

\begin{figure}[H]
\centering
\includegraphics[scale=0.45]{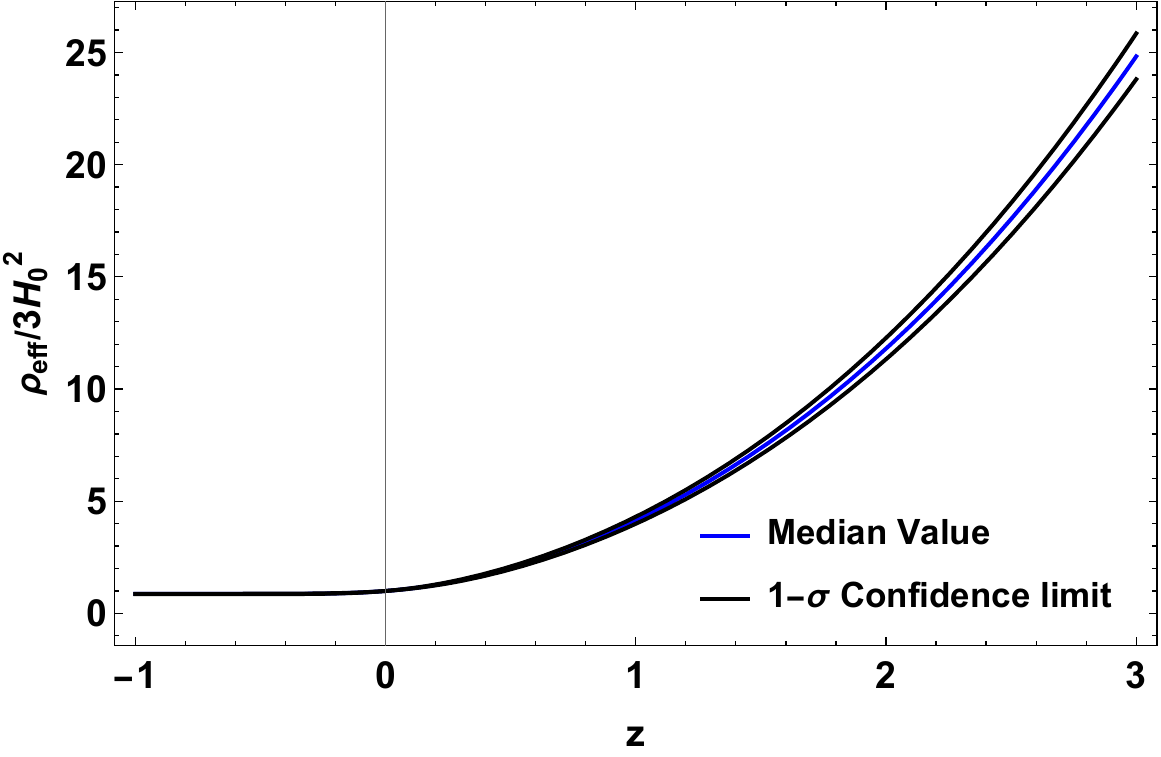}
\caption{Profile of the effective energy density vs redshift corresponding to obtained parameter constraints with $68 \%$ confidence limit.}
\label{f3}
\end{figure}
From Figure \eqref{f2}, it is evident that the assumed model shows a transition from decelerated epoch to the de-Sitter type accelerated expansion phase, with the transition redshift $z_t=0.288^{+0.031}_{-0.029}$. The present value of the deceleration parameter obtained as $q(z=0)=q_0=-0.56^{+0.04}_{-0.03}$ ($68 \%$ confidence limit), that is quite consistent with observed ones. Further from Figure \eqref{f3}, we obtained expected positive behavior of the effective energy density that is vanishes with the expansion of the universe.

\subsection{Thermodynamical Stability}
\justifying
To conduct a thorough assessment, we investigate the thermodynamical stability of the assumed theoretical model by examining the sound speed parameter. In this analysis, we assume that the universe operates as an adiabatic system, where there is no transfer of heat or mass from within the universe to its external environment, resulting in a zero entropy perturbation. Under these conditions, the variation of pressure in relation to energy density becomes the primary focus, leading us to introduce the sound speed parameter, denoted as $c_s^2$, in the subsequent expression,
\begin{equation}\label{4j}
c_s^2=\frac{\partial p}{\partial \rho} =  \frac{\partial_z p}{\partial_z \rho }
\end{equation}
where $\partial_z= \frac{\partial}{\partial z}$. Here, it is noteworthy that the condition $c_s^2 > 0$ indicate stability, while $c_s^2 < 0$ signifies instability.
\begin{figure}[H]
\centering
\includegraphics[scale=0.51]{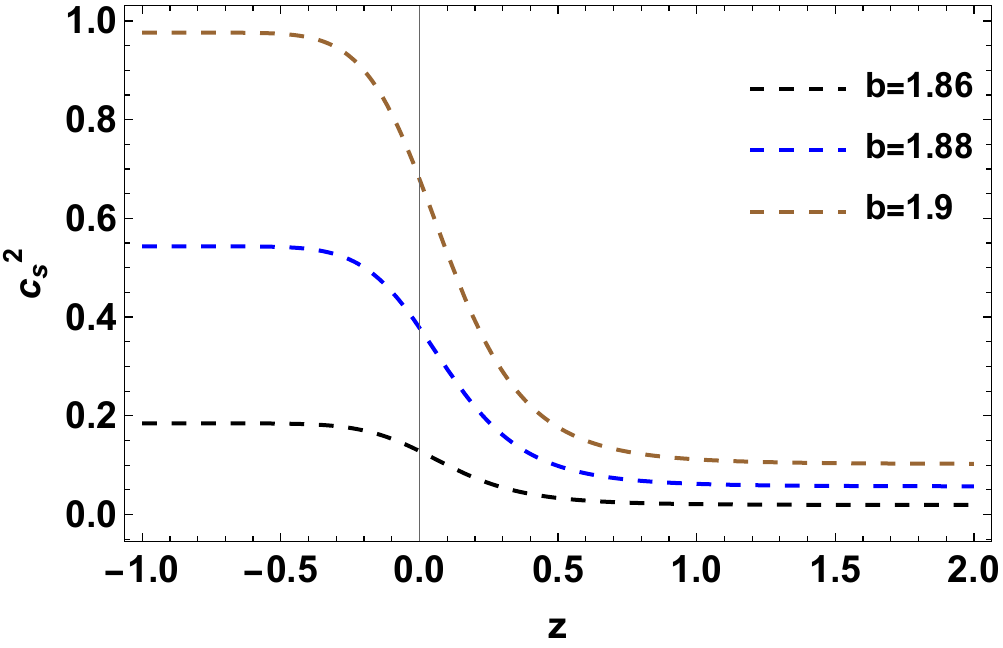}
\caption{Profile of the sound speed parameter vs redshift corresponding to the value $b=1.86$, $b=1.88$, and $b=1.9$. }
\label{f4}
\end{figure}
From Figure \eqref{f4}, it is evident that the assumed theoretical $f(Q)$ model shows the evolution of the universe from decelerated to an accelerated epoch in a stable way. Thus the considered model can efficiently address the late-time expansion phase with the observed transition epoch. 

\section{Conclusion}\label{sec5}
\justifying
In this article, we attempted to explore the dark sector of the universe i.e. dark matter and dark energy. We considered an extended form of the equation of state (EoS) for dark matter, widely known as the Extended Bose-Einstein Condensation (EBEC) EoS for dark matter (presented in the equation \eqref{3c}), with the modified $f(Q)$ lagrangian. The state of matter known as Bose-Einstein condensate (BEC) arises in the non-baryonic realm, when particles called bosons undergo cooling to near absolute zero \cite{R5}. The assumption is made that dark matter exists in the form of a bosonic gas below a critical temperature, leading to the formation of Bose-Einstein condensate (BEC). Utilizing the generalized Gross-Pitaevskii equation, the equation of state (EoS) for dark matter is derived as that of a barotropic fluid. Further, the dark matter halo existing in a quantum ground state, the equation of state (EoS) was derived as $p \propto \rho^2$ \cite{R10}. These observations motivate us to consider the Extended Bose-Einstein Condensate (EBEC) model, a comprehensive model combining normal dark matter and the quantum ground state \cite{R13}.

Now to describe another prominent dark component i.e. undetected dark energy, we consider the modified theories of gravitation within a flat spacetime geometry, dependent solely on non-metricity, particularly, we consider the power law $f(Q)$ lagrangian $f(Q)= \gamma \left(\frac{Q}{Q_0}\right)^n$, where $\gamma$ and $n$ are free parameters \cite{R44}. We present the corresponding Friedmann-like equations and the continuity equation for both dark components along with an interacting term. The interaction term is directly proportional to the product of the Hubble parameter and the energy density of dark matter. In other words, it signifies the energy exchange between the dark sector of the universe. We obtained the analytical solution of the corresponding equations, i.e. the Hubble function in terms of redshift,  presented in the equation \eqref{3i}. Further, to find the best-fit values of parameters of the assumed theoretical model, we utilize the Bayesian analysis to estimate the posterior probability through the utilization of the likelihood function and the Markov Chain Monte Carlo (MCMC) sampling technique. The corresponding contour plot describing the correlation of model parameters within the $1\sigma-3\sigma$ confidence interval utilizing CC+Pantheon+SH0ES samples is presented in Figure \eqref{f1}. The obtained constraints on the free parameter space with $68 \%$ confidence limit are $H_0=72^{+0.11}_{-0.12}$, $n=-2.9^{+0.066}_{-0.067}$, $\beta=0.83^{+0.12}_{-0.11}$, $\eta=-2.5^{+0.06}_{-0.061}$, and $c=0.17^{+0067}_{-0.0068}$. In addition, to examine the robustness of our MCMC analysis, we perform a statistical assessment using the Akaike Information Criterion (AIC) and Bayesian Information Criterion (BIC). We obtained $\Delta AIC=0.85$ and $\Delta BIC=15.5$, and hence it is evident from the $\Delta AIC$ value that there is strong evidence in favor of the assumed theoretical $f(Q)$ model. However, it is well known that a large number of parameters compensate for a high $\Delta BIC $ value. We presented the evolutionary profile of the deceleration parameter and the energy density, respectively in Figures \eqref{f2} and \eqref{f3}. We found that the assumed model shows a transition from the decelerated epoch to the de-Sitter type accelerated expansion phase, with the transition redshift $z_t=0.288^{+0.031}_{-0.029}$. Moreover, the present value of the deceleration parameter obtained as $q(z=0)=q_0=-0.56^{+0.04}_{-0.03}$ ($68 \%$ confidence limit), which is quite consistent with cosmological observations. Further, we found the expected positive behavior of the effective energy density. More on energy density, we obtained $\Omega_0 = \frac{(1-2n)\gamma}{Q_0} $ utilizing the relation \eqref{3f}. For the STEGR case i.e. $n=1$ and $\frac{\gamma}{Q_0}=-1$, we obtained expected value $\Omega_0=1$ that represents matter dominated phase. Further, for the parameter value $n=-2.9$ (obtained in Figure \eqref{f1}), we acquired $\Omega_0 = \frac{6.8 \gamma}{Q_0} $ which aligns with the observed value $\Omega_0 \in [0.25, 0.35] $ for the viable parameter range $\frac{\gamma}{Q_0} \in [19, 27] $. Lastly, we investigated the thermodynamical stability of the assumed theoretical model by examining the sound speed parameter (presented in Figure \eqref{f4}). We observed that one can analyze the contribution of the normal dark matter and the dark matter halo existing in a quantum ground state separately by estimating the value of parameter $\alpha$ using the specific value of intensity parameter $b^2$ and the obtained constrained value of $\eta$, for instance, if we choose $b=1.9$ (obtained in Figure \eqref{f4}) and $\eta=-2.5$ (obtained in Figure \eqref{f1}), we obtained $\alpha=0.11$ utilizing the relation $\eta=\alpha+1-b^2$. As we already have $\beta=0.83$ (obtained in Figure \eqref{f1}), we found that the dark matter halo existing in a quantum ground state contributes nearly $7.5$ times more than that of the normal dark matter.we found that the considered theoretical $f(Q)$ model can efficiently address the late-time expansion phase of the universe with the observed transition epoch in a stable way.

\textbf{Data availability:} There are no new data associated with this article.\\

\section*{Acknowledgments} \label{sec7}
 Aaqid Bhat expresses gratitude to the BITS-Pilani, Hyderabad campus, India, for granting him a Junior Research Fellowship. RS acknowledges UGC, New Delhi, India for providing Senior Research Fellowship with (UGC-Ref. No.: 191620096030). PKS acknowledges Science and Engineering Research Board, Department of Science and Technology, Government of India for financial support to carry out Research project No.: CRG/2022/001847 and IUCAA, Pune, India for providing support through the visiting Associateship program. We are very much grateful to the honorable referee and to the editor for the illuminating suggestions that have significantly improved our work in terms
of research quality, and presentation.

%%%%%%%%%%%%%%%%%%%%%%%%%%%%%%%%%%%%%%%%%%%%%%%%%%%%%%%%%%%%%%%%%%%%%%%%%%%%%%%%%%
%%%
%%%

\end{document}